\documentclass[doublecol]{epl2} 
\usepackage{amsmath}
\usepackage{amsfonts}
\usepackage{amssymb}
\usepackage[T1]{fontenc}
\usepackage{subfig}
\usepackage[percent]{overpic}
\usepackage{float}

\newcommand{\be}{\begin{equation}}
\newcommand{\ee}{\end{equation}}
\newcommand{\bc}{\begin{center}}
\newcommand{\ec}{\end{center}}
\newcommand{\bi}{\begin{itemize}}
\newcommand{\ei}{\end{itemize}}
\newcommand{\ba}{\begin{eqnarray}}
\newcommand{\ea}{\end{eqnarray}}

\newcommand{\ignore}[1]{}

\title{Effect of degree correlations above the first shell on the percolation transition}

\author{L. D. Valdez\inst{1}\thanks{E-mail:
    \email{ldvaldes@mdp.edu.ar}} \and C. Buono\inst{1} \and
  L. A. Braunstein\inst{1,2} \and P. A. Macri\inst{1}}

\institute{ 
\inst{1} Instituto de Investigaciones F\'isicas de Mar del Plata
  (IFIMAR)-Departamento de F\'isica, Facultad de Ciencias Exactas y
  Naturales, Universidad Nacional de Mar del Plata-CONICET, Funes
  3350, (7600) Mar del Plata, Argentina.\\ 
\inst{2} Center for Polymer
  Studies, Boston University, Boston, Massachusetts 02215, USA 
}

\shortauthor{L. D. Valdez \etal}


\abstract{The use of degree-degree correlations to model realistic
  networks which are characterized by their Pearson's coefficient, has
  become widespread. However the effect on how different correlation
  algorithms produce different results on processes on top of them,
  has not yet been discussed.  In this letter, using different
  correlation algorithms to generate assortative networks, we show
  that for very assortative networks the behavior of the main
  observables in percolation processes depends on the algorithm used
  to build the network. The different alghoritms used here introduce
  different inner structures that are missed in Pearson's
  coefficient.  We explain the different behaviors through a
  generalization of Pearson's coefficient that allows to study the
  correlations at chemical distances $\ell$ from a root node.  We
  apply our findings to real networks.}

\begin{document}
\maketitle

\section{Introduction}
In the last two decades, the use of complex networks in the study of
many processes, such as the spread of diseases, random or intentional
attacks, synchronization, etc.~\cite{Boc_01,Dor_02,Dor_03}, has led to a deeper
understanding of these processes. For example, the use of
network-based models in epidemiology has demonstrated how the topology
affects the total fraction of the infected population and how it's
knowledge can be used to develop efficient immunization
strategies~\cite{Coh_03}.  Each of the aspects of the network topology
has a substantial effect on the processes taking place on the
underlying network.

A first feature of the network topology is the degree distribution
$P_{d}(k)$, i.e., the fraction of nodes with degree $k$. One of the
most used networks is the Erd\"os R\'enyi (ER)~\cite{Erd_01} with
$P_{d}(k)=e^{-\langle k\rangle}\langle k\rangle ^{k}/k!$, where
$\langle k\rangle$ is the average degree. However in real networks the
degree distribution fits better to a scale-free (SF) $P_{d}(k)\sim
k^{-\lambda}$, where $\lambda$ is the broadness of the
distribution. Many of these networks also have an exponential cutoff
$\kappa$~\cite{Ama_01}, and the degree distribution can be represented
by $P_{d}(k)\sim k^{-\lambda}e^{-k/\kappa}$.

Several analytical and numerical models on degree-degree uncorrelated
networks were developed with the above
distributions~\cite{New_03,Mil_02}. However, in the last years it has
been observed that in real networks, the degrees of the linked nodes
are correlated. For example, it is known that in social networks,
nodes tend to be linked to others with similar degree, while in
technological networks, large degree nodes tend to be connected with
low-degree ones. Networks with the first pattern of connections are
called assortative, while those with the second pattern are
disassortative.  Recent researches suggest that networks adopt a
correlated structure in order to optimize some processes that are
developed on top of them, such as synchronization~\cite{Sor_01},
transport~\cite{Yu_01}, traffic dynamic~\cite{Jin_01},
congestion~\cite{Ana_01} and growth~\cite{Jia_01}. As a consequence,
it is necessary to consider models of networks with degree-degree
correlations in order to study how the processes are affected by
them. There are different measures to quantify the correlations
between nearest neighbors~\cite{Boc_01,Dor_02}.  The most detailed
measure is the joint degree distribution $P_{e}(k,k')$, which
indicates the probability that a node of degree $k$ is linked with
another of degree $k'$. However, due to the large amount of
information contained in this distribution, it is difficult to
interpret the results based on $P_{e}(k, k ')$. Hence more global
measures are needed to simplify the interpretation of the results.
For example, a measure which follows from the above and is more
simplified is the average nearest neighbor degree of a node with
connectivity $k$, $k_{nn}(k)=\sum_{k'}k'P_{e}(k|k')$. The sign of the
slope of $k_{nn}(k)$ is positive (negative) for assortative
(disassortative) networks. However, the most used measurement is
Pearson's coefficient, which is less detailed than the previous ones
and gives a general overview of the correlation to nearest neighbors
on the network. Pearson's coefficient is given by

\begin{equation}\label{Eq.r}
r= \frac{\langle k k'\rangle_{e} - \langle (k + k^{'})/2\rangle^{2}_{e}}{\langle
 (k^2 + k^{'2})/2\rangle_e- \langle (k + k^{'})/2\rangle^{2}_{e} }\;,
\end{equation}
where $e$ is the average over $P_{e}(k,k')$. Assortative networks have
$r> 0$, while disassortative networks have $r <0$. Finally, networks
with $r = 0$ are called uncorrelated.

Many studies that concentrate on how the degree-degree correlation
affects the processes on top of networks, use Pearson's
coefficient to quantify the results. For example in Ref.~\cite{Sor_01}
it has been observed that synchronization is enhanced for $r\neq
0$. Also, in Ref.~\cite{Ana_01}, under a gradient network formalism,
the authors explain why in disassortative networks the congestion is
lower than in uncorrelated and assortative ones. On the other hand,
some studies on the effect of degree-degree correlations were focused
on the link percolation process~\cite{Vaz_01,Xul_01,Mor_01,Hoo_01}. The
investigations on this process are very useful since they provide
information about how resistant a network is against random failures
and also due to the relation between percolation with some aspects of
the spread of diseases~\cite{Dun_01,Coh_01,Coh_02,New_05,Gra_01}.  For
uncorrelated networks, it is well known that in percolation processes
there is a critical fraction $p=p_{c}$ of nodes/links, above which the
network undergoes a second order phase
transition~\cite{New_03,Coh_01}. The order parameter of the transition
is $P_{\infty}(p)$, which is the fraction of nodes that belong to the
largest component for a given value of $p$.  For infinite systems
($N\to \infty$) and close to the threshold ($p_{c}$), the order
parameter behaves as~\cite{Sta_01}
\begin{equation}\label{Eq.pinf}
 P_{\infty}(p) \sim (p -p_c)^\beta, \,\,\,\,\,\,  p\gtrsim p_{c}+\delta p,\,\,\delta p>0;
 \end{equation}
and the average value of the size of the finite clusters $\langle
s(p)\rangle$ goes as
 \begin{equation}\label{Eq.S}
  \langle s(p)\rangle \sim (p -p_c)^{-\gamma}, \,\,\,\,\,\, p\lesssim p_{c}-\delta p,\,\,\delta p>0;
 \end{equation}
where $\beta=1$ and $\gamma=1$ in the mean field
approach~\cite{Sta_01}.  In the case of finite networks with $N$
nodes, Eq.~(\ref{Eq.S}) obey the general scaling law
 \begin{equation}\label{Eq.S1}
   \langle s \rangle \equiv \langle s (p,N)\rangle =N^{\gamma \; \Theta}f\Bigl(\bigl(p-p_{c}(N)\bigr)N^{\Theta}\Bigr),
 \end{equation}
where $f$ is a scaling function that behaves as a constant at the
threshold, with $\langle s \rangle \sim N^{1/3}$ for
uncorrelated ER and SF networks with $\lambda \geq
4$~\cite{Han_01,Wu_01}.

 Newman~\cite{New_01} and Miller~\cite{Mil_01} showed that networks
 with $r>0$ are more resilient to damage than those with $r\leq
 0$. However, there has been little discussion in the literature about
 whether the percolation phase transition remains of second order
 depending on the value of $r$.  This issue was only addressed in
 Ref.~\cite{Noh_01} where they found that in correlated networks
 constructed using a Metropolis algorithm and characterized by $r$,
 for the disassortative case percolation belong to the same
 universality class as uncorrelated networks but highly assortative
 networks have no second order phase transition. A recent
 investigation by Dorotgsev {\it{et. al}}~\cite{Dor_01} pointed out
 that recursive trees can be constructed with a highly correlated
 structure but with zero Pearson's coefficient. On the other hand, in
 Ref.~\cite{Noh_02} the authors show that positive correlations
 affects the loop statistics on networks. Thus, Pearson's
 coefficient could hide a more complex pattern of connections between
 nodes that also affect the processes that spread on networks. This
 raises the question as to whether or not Pearson's coefficient is a
 reliable tool for characterizing correlated networks
 and, in particular, the percolation transition on assortative
 networks.

In this letter we use different correlation algorithms to determine
the effect of very assortative networks on the link percolation
transition. We find that the results depend strongly on the algorithm
used to correlate the networks and not on the value of Pearson's
coefficient.  Moreover, we find that strong correlations above the
first shell on assortative networks affect the percolation transition.

\section{Correlation Algorithms}\label{modelos}

We will focus only on assortative networks because we want to
understand the effect of different algorithms in the percolation
transition.

In this work we use rewiring algorithms to correlate the networks,
since they preserve the degree distribution $P_{d}(k)$. In particular
we focus on two correlation algorithms: the Exponential Random Graph
(ERG)~\cite{Noh_01} and the Local Optimal Algorithm (LOA) introduced by us.

For the ERG algorithm, the process to correlate a network or graph
$G$~\cite{Noh_01}, is based on a Metropolis dynamic, which uses a
``Hamiltonian'' function given by $H(G)=-J
\sum_{i,j>i}A_{ij}k_{i}k_{j}$, where $A_{ij}$ is the component of the
adjacency matrix, $k_i$ ($k_j$) is the connectivity of node $i$ ($j$)
and $J$ is a control parameter for the correlation.  For $J>0$ ($J<0$)
an assortative (disassortative) network is built. This algorithm
generates a Gibbsian ensemble network and is ergodic. The algorithm
correlates the network, successively applying the following steps:
\begin{enumerate}
\item in a network $G$, two links are chosen randomly to be correlated.
\item rewiring is allowed (from configuration $G$ to $G'$) with
  probability $w=\min\{1,\exp[-(H(G')-H(G))]\}$, disallowing autoloops
  and multiple connections.
\end{enumerate}
The rewiring process continues until the steady state, where for large
enough system sizes, $r$ depends only on $J$ (not shown here).

In the LOA algorithm, a network is correlated as follows:

\begin{enumerate}

\item two links are randomly chosen from the network to be correlated.

\item for assortative (disassortative) networks, we choose the
  configuration that increases (decreases) most Pearson's coefficient
  $r$ without generating autoloops or multiple connections.
\end{enumerate}
This algorithm is not ergodic but achieves the desired
correlation faster, since in contrast to the previous one, we do not have to
wait for the equilibrium value in order to reach the desired value of
$r$.

\section{Results and Discussion}

\subsection{Link percolation on assortative networks}

In this section we study the effects of the correlation algorithms on
the percolation transition in assortative networks.

\begin{figure}[h]
\vspace{0.3cm}
\centering
\includegraphics[scale=0.25]{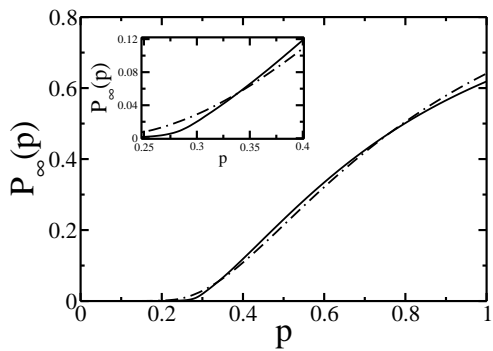}
\caption{Plot of $P_{\infty}(p)$ for ER networks with $\langle k \rangle =2$,
  $N=8\times 10^4$ and $r=0.78$, correlated with LOA (solid line) and ERG
  (dashed line, $J=1$). In the inset we show an enlargement of the main plot close
  to the threshold. The simulations were done over $10^{3}$
  realizations.\label{fig.Pinf}}
\end{figure}

In Fig.\ref{fig.Pinf}, we show $P_{\infty}(p)$ for an ER network with
fixed $N$ and $r$ for the ERG and LOA models.  Even though the curves do
not differ significantly between them, in the inset we can see that
$P_{\infty}(p)$, close to the threshold computed with different
algorithms converge to zero with different slopes, suggesting that the
transitions are different even if the same value of $r$ is used in
both algorithms. The behavior of $P_{\infty}(p)$ for networks
generated with the ERG algorithm seems to be nonsingular.

\begin{figure}[h]

\centering
\vspace{0.3cm}
  \begin{overpic}[scale=0.25]{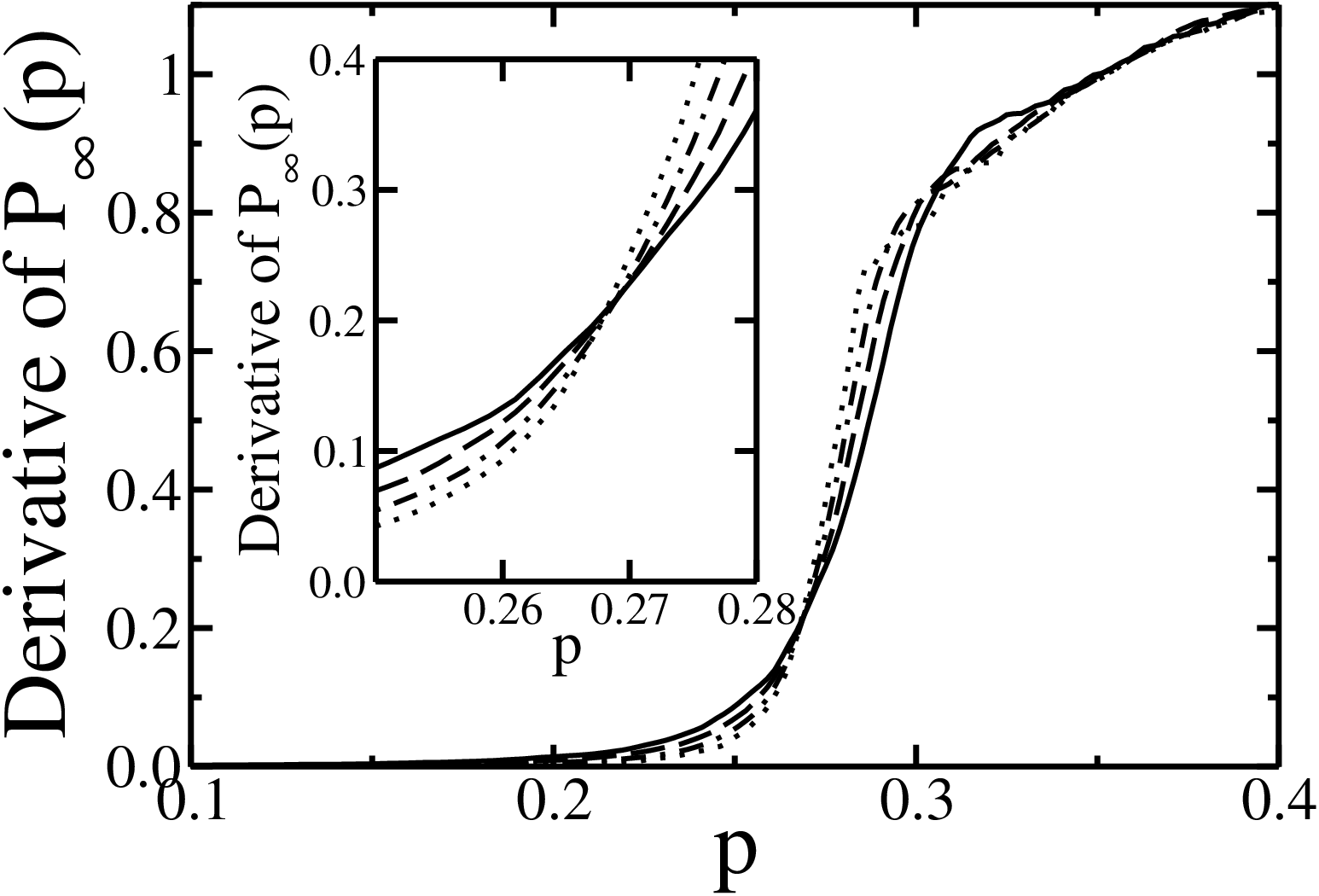}
    \put(80,20){{\bf{(a)}}}
  \end{overpic}\vspace{1cm}
  \begin{overpic}[scale=0.25]{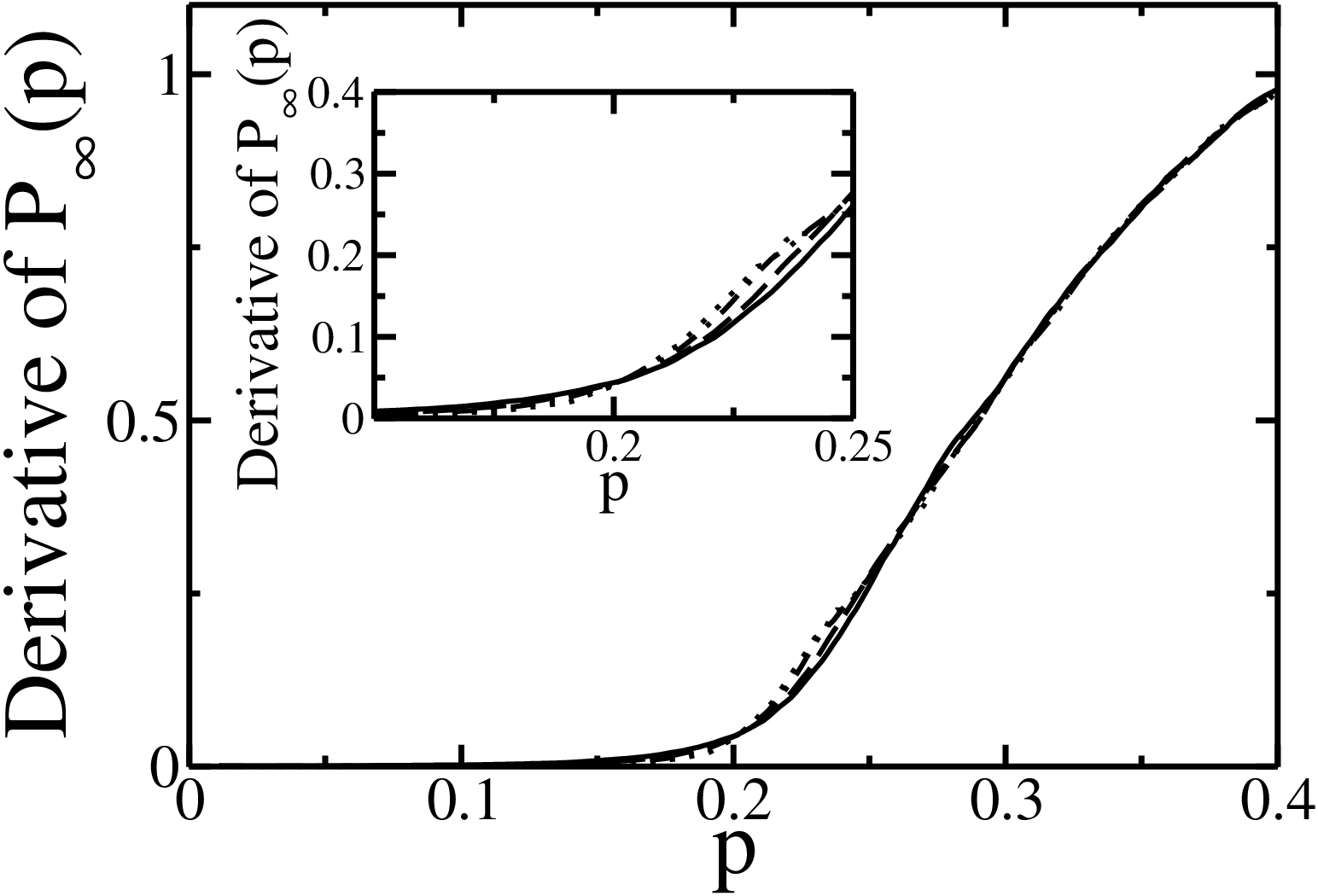}
    \put(80,20){\bf{(b)}}
  \end{overpic}\vspace{0.5cm}
\caption{Plot of the derivative of $P_{\infty}$ as a function of $p$
  for ER with $\langle k \rangle =2$, $r=0.78$, correlated with LOA
  (a) and ERG (b) for: $N=8\times 10^4$ (solid line), $N=16\times
  10^4$ (dashed line), $N=32\times 10^4$ (dash-dotted line) and
  $N=64\times 10^4$ (dotted line). In the insets we show an
  enlargement of the main plots close to the threshold. The
  simulations were done over $10^3$
  realizations. \label{fig.deriv_Pinf}}
\end{figure}

It is well known that in the thermodynamic limit a second order phase
transition, has a singularity in the
derivative of the order parameter at the threshold $p_{c}$. In
Fig.\ref{fig.deriv_Pinf} we plot the derivative of $P_{\infty}(p)$
with respect to $p$ for different networks sizes. From the plots, we
can see that in strong positive correlated networks with the LOA algorithm, the
derivative of $P_{\infty}(p)$ exhibit a singularity as we increase the
system size, characteristic of a second order phase
transition. However for the ERG model the slope is smooth and does not
depend on $N$. Therefore, we confirm that the transitions for the ERG and
the LOA are not the same~\cite{Noh_01}.

\begin{figure}[h]
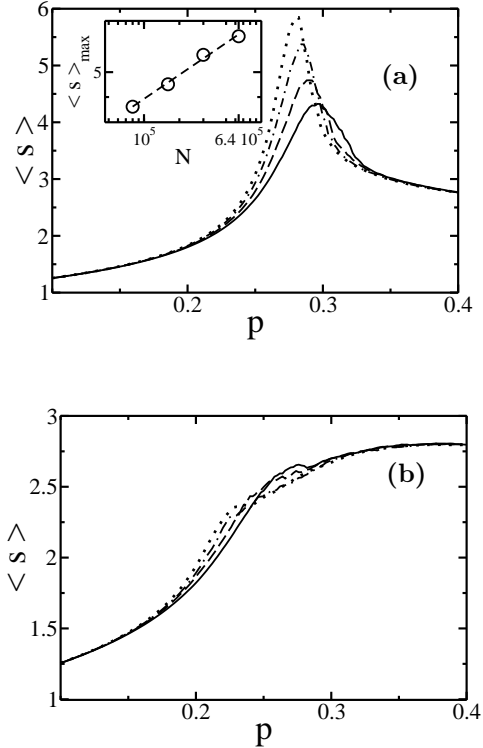

\vspace{0.3cm}
\centering
  \begin{overpic}[scale=0.25]{Fig_04.eps}
    \put(80,56){\bf{(a)}}
  \end{overpic}\vspace{0.9cm}
  \begin{overpic}[scale=0.25]{Fig_05.eps}
    \put(80,56){\bf{(b)}}
  \end{overpic}\vspace{0.5cm}
\caption{Plot of $\langle s \rangle$ as a function o $p$ for the same
  networks as in Fig.\ref{fig.deriv_Pinf}: LOA (a) and ERG (b). In the
  inset of (a) we plot the height of the peak of $\langle s \rangle$ as
  a function of $N$ on a log-log scale. The dashed line represents a
  power law fitting from which we obtain
  $\gamma\Theta=0.16$ (see Eq.~(\ref{Eq.S1})).\label{fig.smed}}
\end{figure}

In order to verify the existence of a second order phase transition,
we compute $\langle s \rangle$ for ER and different network sizes as
shown in Fig.\ref{fig.smed}. We can see that with the LOA correlation
model as $N$ increases, the peak of $\langle s \rangle$ increases
around the critical threshold $p_{c}(N)$ as in a second-order phase
transition~\cite{Sta_01}. At $p=p_{c}$ (see Eq.~(\ref{Eq.S1}))
$\langle s \rangle \sim N^{0.16}$, suggesting that percolation in
assortative networks generated by the LOA belong to a different
universality class from that in uncorrelated networks. As noted by Noh
using the ERG model~\cite{Noh_01}, $\langle s \rangle$ has no peak
independent of the value of $N$, which still indicates that the
transition is not of second order.

\begin{figure}[h]
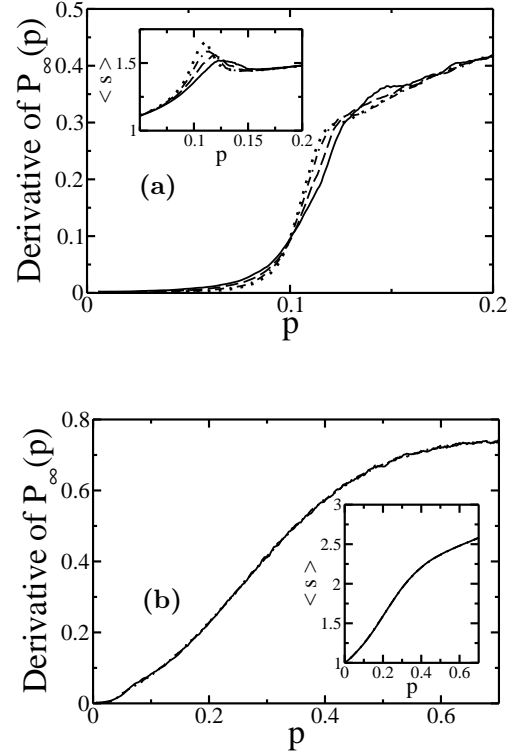

\vspace{0.3cm}
\centering
  \begin{overpic}[scale=0.25]{Fig_06.eps}
    \put(25,30){\bf(a)}
  \end{overpic}\vspace{0.9cm}
  \begin{overpic}[scale=0.25]{Fig_07.eps}
    \put(25,30){\bf(b)}
  \end{overpic}\vspace{0.5cm}
\caption{Derivative of $P_{\infty}(p)$ for SF networks with
  $\lambda=2$, $\kappa=10$ and $r=0.55$ correlated with LOA (a) and
  ERG (b) for: $N=8\times 10^4$ (solid line), $N=16\times 10^4$
  (dashed line), $N=32\times 10^4$ (dash-dotted line) and $N=64\times
  10^4$ (dotted line). In the inset we plot $\langle s \rangle$ as a
  function of $p$ for LOA (a) and ERG (b).\label{SF_trans}}
\end{figure}

We run all our simulations for SF networks with exponential cutoff and
find similar behavior that for ER networks. In Fig.~\ref{SF_trans} we
show that the derivative of $P_{\infty}(p)$ and $\langle s \rangle$ do
not exhibit any singularity as we increase the network size for the ERG
model, in contrast to the LOA model.

Despite the discrepancies between the assortative networks, we find
that disassortative networks generated by both algorithms undergo a
second-order phase transition with $\langle s \rangle \sim N^{1/3}$,
which not only confirms that disassortative networks belong to the
same universality class as uncorrelated ones, but also that the
critical exponents are independent of the algorithm used to generate
negative correlations. Consequently, we conclude that while
Pearson's coefficient is a good measure to uniquely characterize the
percolation transition on disassortatives and uncorrelated networks,
it fails to characterize the transition on very assortative networks.
The discrepancies between the assortative networks generated with
different algorithms must underlie on structural differences
introduced by the algorithms, which cannot be explained by only
Pearson's coefficient. Hence, we will show below the different effects
that both algorithms introduce on the network topology.

\subsection{Correlations above the first shell}

In order to explain the differences between assortative networks with
the same value of $r$ but correlated with different algorithms, we
propose a new measure called generalized Pearson's coefficient
$r_{\ell}$, which measures the correlation between the connectivity of
a root node and the neighbors at a chemical distance $\ell$, i.e. which are
located in the $\ell-th$ shell from the root . In this way, we denote the
usual Pearson's coefficient $r$ as $r_{1}$, Pearson's coefficient
to second neighbors as $r_{2}$ and so forth. We define,

\begin{equation}\label{Eq.rl}
r_{\ell}= \frac{\langle k k'\rangle_{\ell} - \langle (k + k^{'})/2\rangle^{2}_{\ell}}{\langle
 (k^2 + k^{'2})/2\rangle_{\ell}- \langle (k + k^{'})/2\rangle^{2}_{\ell} }\;,
\end{equation}
where $\langle \ldots \rangle_{\ell}$ is the average over
$P_{\ell}(k,k'_{\ell})$, where $P_{\ell}(k,k'_{\ell})$ is the
probability that a node of degree $k$ is connected to a node of degree
$k'$ at the $\ell-th$-shell.

\begin{figure}[h]
\vspace{0.1cm}
\centering
\includegraphics[scale=0.25]{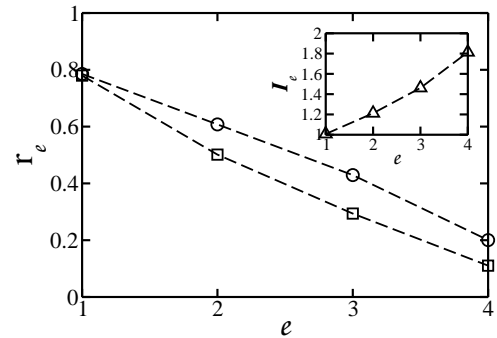}
\caption{Plot of $r_{\ell}$ as a function of the shell number $\ell$
  for LOA ($\square$) and ERG ($\bigcirc$) in ER networks with
  $\langle k \rangle =2$, $r=r_{1}=0.78$ and $N=10^{4}$. The dashed
  lines are used only as a guide to the eye. In the inset, we show
  $I_{\ell}$ as a function of the shell number $\ell$. As seen in the
  plot $I_{2}=1.21$, $I_{3}=1.46$, $I_{4}=1.81$. Then the ERG model
  generate stronger long range correlations than the LOA model. The
  simulation were done over $10^4$ realizations.\label{fig.r_elles}}
\end{figure}

In Fig.\ref{fig.r_elles} we show for ER networks $r_{\ell}$ as a
function of $\ell$ in assortative networks~\footnote{The understanding
  of the disassortative behavior of $r_{\ell}$ goes beyond the scope
  of this work.}, for $r_{\ell}>0$. We can see that $r_{\ell}$ is a
decreasing function of $\ell$. Imposing high assortativity in the
network, correlations above the first shell from a root node are
generated. Thus, correlation will build groups of nodes connected with
similar degrees at shells close to the root, forming groups of nodes
of low degrees and others with high degree. The high degree groups
will form a strong core very resilient to random failures. The
stronger the correlations are above the first shell, the more
resilient groups are, smoothing out the transition.

In the inset of Fig.\ref{fig.r_elles} we plot the factor $I_{\ell}$
defined as the ratio between $r_{\ell}$ for ERG and $r_{\ell}$ for LOA
networks. Then $I_{\ell}>1$ ($I_{\ell}<1$) indicates that ERG networks
have higher (lower) degree correlation at a distance $\ell$ than LOA
networks. It is easy to see that for shells above $\ell=1$, $I_{\ell}$
increases with $\ell$ reaching a value of $I_{4}=1.8$ for the fourth
shell\footnote{For a SF network with $\lambda=2$, $\kappa=10$,
  $r=0.55$ and $N=10^4$, we obtain that $I_{2}= 2.11$, however
  positive correlations only reach to the second shell.}  for
$N=10^{4}$. Thus ERG networks generate stronger correlation above
shell $\ell=1$ than LOA networks supporting our picture that high
assortative ERG networks posses a stronger structure of groups, with
nodes of similar connectivities, than in the LOA model. This
introduces two aspects of the network that contributes to erase the
second-order phase transition in the ERG model. First, ERG networks
have more homogeneous groups of nodes with low connectivity that are
easier to fragment, generating smaller clusters than in LOA
assortative networks.  On the other hand, in ERG networks there are
also more homogeneous groups of highly connected nodes that remain in
the largest component through the percolation
process\footnote{The same picture can be seen in onion
    networks~\cite{Sch_01,Wu_02}. However onion and assortativity are
    distinct properties as pointed out in \cite{Sch_01}.}, preventing
the network from fragmenting into large finite clusters, and smoothing
out the curve of $P_{\infty}(p)$ near the threshold (see
Fig.\ref{fig.deriv_Pinf}). Consequently, these two aspects induce a
greater presence of small clusters, leading to a non-diverging
$\langle s \rangle$.
\begin{figure}[h]
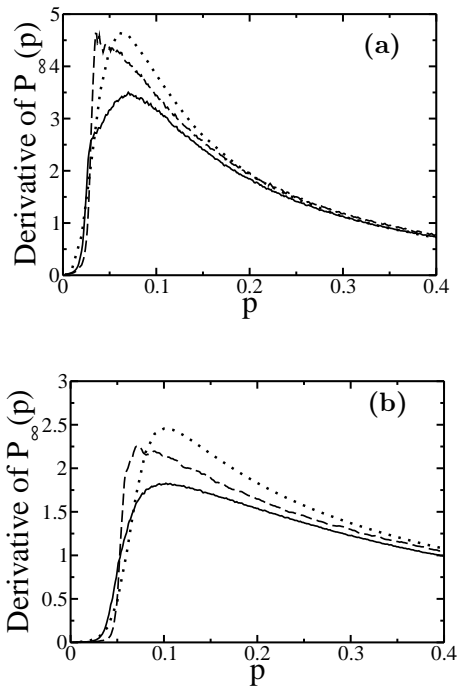

\vspace{0.1cm}
\centering
  \begin{overpic}[scale=0.23]{Fig_09.eps}
    \put(80,61){\bf(a)}
  \end{overpic}\vspace{0.7cm}
  \begin{overpic}[scale=0.23]{Fig_10.eps}
    \put(80,61){\bf(b)}
  \end{overpic}\vspace{0.1cm}
\caption{Plot of the derivative of $P_{\infty}$ as a function of $p$
  close to the threshold for a) the cond-mat coauthorship network with
  $N=40421$ and $r=0.18$ and b) the mathematics coauthorship network with
  $N=391529$ and $r=0.12$. The solid lines correspond to percolation
  on the original data, and dashed and dotted lines correspond to LOA
  and ERG networks, respectively with the same degree distribution as
  the real data. We can see that all the curves exhibit a pronounced
  steep growth at the threshold characteristic of a second order phase
  transition.\label{fig.deriv_reales}}
\end{figure}
 To determine if this behavior is representative of real networks, we
 measure $r_{\ell}$ on two different positive correlated SF real
 graphs, the condensed matter (cond-mat, $\lambda\approx 1.60$ and
 $\kappa\approx 40$)~\cite{New_04} and mathematics ($\lambda \approx
 1.72$ and $\kappa\approx 35$)~\cite{Pal_01} coauthorship networks. In
 both cases, we obtain that positive correlations are non-negative
 only until the second shell with $r_{1}=r=0.18$, $r_{2}=0.01$ for the
 cond-mat collaboration network and $r_{1}=r=0.12$, $r_{2}=0.03$ for
 the mathematics collaboration network. Consequently, real networks
 posses mainly correlation to first neighbors ($r$), in contrast to
 the stronger correlation structure above $\ell=1$ found in the
 theoretical networks, particularly for ERG, and therefore the real
 graphs analyzed here have a less defined structure of groups of nodes
 connected with similar degrees, suggesting that there are constrains
 which prevent networks from evolving towards extreme
 correlations. This can be seen, for example, on both coauthorship
 networks used here. In general, coauthorship networks are made of
 research groups composed by senior researchers with high connectivity
 and young researchers generally with low connectivity. On one hand,
 senior researchers from different research groups have some
 connections between themselves in order to increase the resources and
 importance of their groups. Moreover, young researchers from the same
 group collaborate between themselves, because they generally work on
 similar projects. This explains the assortativity of the networks. On
 the other hand, in order to share their knowledges, senior
 researchers are highly connected to the young ones of their group,
 decreasing the assortativity. Thus the network evolves with an
 assortativity correlation, but with a not too high Pearson's
 coefficient in order to improve the functionality of the full
 coauthorship network. This is why real assortative collaboration
 networks are not extreme assortativity correlated. Moreover, since
 this positive correlation decreases with $\ell$ as we can see in both
 real networks used here and in Router graph studied in
 Ref.~\cite{Ech_01}, the correlation above the first shell is almost
 zero due to the small Pearson's coefficient, which leads to a
 critical behavior of percolation process on these networks. In
 Fig.\ref{fig.deriv_reales} we plot for the real networks, the
 derivative of $P_{\infty}(p)$ for the original data and after
 applying the LOA and the ERG models for networks with the same degree
 distribution. As we can see, there is a steep growth of the
 derivative near criticality, which is equally sharp in the LOA and
 ERG networks, supporting our argument that in real networks the
 second order phase transition exists. In turn, we note that ERG and LOA
 networks with the same degree distribution and Pearson's coefficient
 as real networks, have also a second order phase percolation
 transition. Our results suggests that Pearson's coefficient is a
 good indicator of the percolation transition order when the network
 is not strongly correlated above the first shell, but for high positive
 degree-degree correlations, different algorithms may generate
 different behaviors in percolation and therefore, Pearson's
 coefficient is no longer useful to indicate the behavior of the
 transition. As a consequence, it becomes crucial to use measures that
 take into account the inner structure of the network, such as
 $r_{\ell}$, that as we show here is a better indicator of whether a
 network will undergo or not a critical phase percolation transition.

In summary, we find that the second order percolation transition on
theoretical networks depends on the inner structure imposed by the
algorithms used to build them, and not only on Pearson's
coefficient. This means that Pearson's coefficient hides a
long-range correlation that could change dramatically the behavior on
top of them. We propose a new magnitude $r_{\ell}$, which allows to
explain the discrepancies between percolation transition on
assortative networks generated by different correlation
algorithms. For the real networks analyzed in this letter, we find
that they posses mainly first neighbors correlation, and thus the
percolation transition is of second order.

\acknowledgments This work was supported by
UNMdP and FONCyT (PICT 0293/2008).  The authors thanks Ana L. Pastore
y Piontti and Joel C. Miller for useful discussions.

\bibliographystyle{eplbib.bst}
\bibliography{bibpaper_corr}

\end{document}